\documentclass[11pt,twoside]{article}  
\usepackage{asp2006}
\usepackage{adassconf}

\begin{document}   

%
%

\paperID{O1.3}

%

\title{The Dark Energy Survey Data Management System: The Processing Framework}

%
%
%
%
%

\markboth{DESDM: The Processing Framework}{Gower, et al.}

%
%
%
%

\author{Michelle Gower, Joseph J. Mohr, Darren Adams, Y. Dora Cai, Gregory E. Daues, Tony Darnell, Chow-Choong Ngeow, Shantanu
Desai, Cristina Beldica, Mike Freemon}
\affil{University of Illinois, Urbana, IL, USA}
\author{Huan Lin, Eric H. Neilsen, Douglas Tucker}
\affil{Fermi National Accelerator Laboratory, Batavia, IL, USA}
\author{Emmanuel Bertin}
\affil{Institut d' Astrophysique, Paris, France}
\author{Luiz A. Nicolaci da Costa, Leandro Martelli, Ricardo L. C. Ogando}
\affil{Observatorio Nacional R. Gal. Jose Cristino, Rio de Janeiro, RJ, Brazil}
\author{Michael Jarvis}
\affil{University of Pennsylvania, Philadelphia, PA, USA}
\author{Erin Sheldon} 
\affil{Brookhaven National Laboratory, New York, NY, USA}

%

\contact{Michelle Gower}
\email{mgower@ncsa.uiuc.edu}

%
%
%

\paindex{Gower, M.}
\aindex{Mohr, J.~J.} 
\aindex{Adams, D.}
\aindex{Cai, Y.~D.}
\aindex{Daues, G.~E.}
\aindex{Darnell, T.}
\aindex{Ngeow, C.} 
\aindex{Desai, S.}
\aindex{Beldica, C.}
\aindex{Freemon, M.}
\aindex{Lin H.}
\aindex{Neilsen E.~H.} 
\aindex{Tucker, D.}
\aindex{Bertin, E.}
\aindex{da Costa, L.~A.~N.}
\aindex{Martelli, L.}
\aindex{Ogando, R.~L.~C.}
\aindex{Jarvis, M.}
\aindex{Sheldon, E.}

%



\begin{abstract}          

The Dark Energy Survey Data Management (DESDM) system will process and
archive the data from the Dark Energy Survey (DES) over the five year
period of operation. This paper focuses on a new adaptable processing
framework developed to perform highly automated, high performance data
parallel processing.  The new processing framework has been used to
process 45 nights of simulated DECam supernova imaging data, and was
extensively used in the DES Data Challenge 4, where it was used to
process thousands of square degrees of simulated DES data.

\end{abstract}

%
%

\section{Introduction}

The Dark Energy Survey (DES, 2011-2016) is an optical survey of
5000 deg\textsuperscript{2} of the South Galactic Cap to \verb+~+24th
magnitude in multiple filter bands (grizY) using a new wide field CCD
camera, DECam, mounted on the Blanco 4-m telescope at Cerro Tololo
Inter-American Observatory (CTIO).  The DECam is a large focal plane
array with a short readout time which will collect approximately 300~GB
of science images per night of observation. Additional data products
will be generated through a series of image processing steps resulting
in a total of approximately 3~TB (uncompressed) of data for each night.

\begin{figure}[t] \epsscale{0.75}\plotone{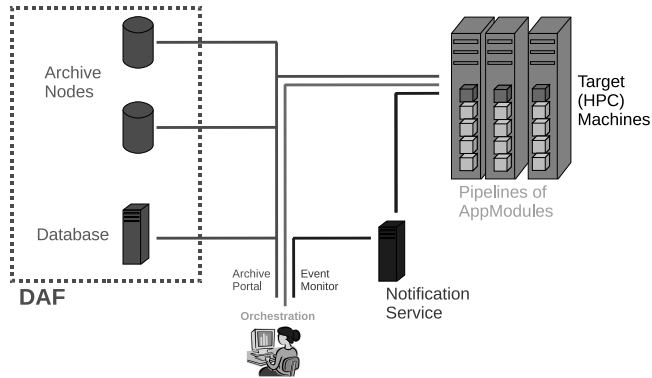} \caption{DESDM Overview.} \label{O1.3_1} \end{figure}

The Dark Energy Survey Data Management system (DESDM\footnote{DESDM
URL: http://desweb.cosmology.uiuc.edu}) 
(see Figure~\ref{O1.3_1})
will process and archive the
data from the Dark Energy Survey (DES) over the five year period of
operation. This paper focuses on a new adaptable processing framework
developed to perform highly automated, high performance data parallel
processing.  The new processing framework has been used to process 45
nights of simulated DECam supernova imaging data, and was extensively
used in the DES Data Challenge 4, where it was used to process thousands
of square degrees of simulated DES data.

\section{Processing Framework}

The processing framework consists of two main components: (1) modular
pipelines that execute science codes and (2) an orchestration layer
to manage the pipelines.  The DESDM pipelines make use of application
containers to wrap astronomical pipeline modules to be executed on
the compute nodes of HPC resources.  These application containers
send events to the Notification service for viewing by the operator.
The orchestration layer of the DESDM processing framework is responsible
for preparing job descriptions (using operator input parameters coupled
with the results of database queries), deploying required files and data
to high-performance computing platforms, and executing and monitoring the
sets of data-parallel jobs that comprise the astronomical processing.
It interfaces with the DESDM Data Access Framework for efficient and
reliable transfer of files and image data to compute platforms.

\subsection{Application Modules}

DESDM processing pipelines are constructed with a Java middleware layer provided by
the Elf/OgreScript\footnote{Elf/OgreScript URL: https://wiki.ncsa.uiuc.edu/display/MRDPUB/Elf} 
software developed at the NCSA. Elf is an robust application container,
and OgreScript is a workflow scripting language with scripts encoded in XML.
The container middleware wraps the astronomical modules to be executed on the 
compute nodes of a target machine, for example, a TeraGrid cluster. 
Elf/OgreScript allows important status
and quality assurance information to be issued within events from running applications.
The container middleware supports the parsing of the streaming stdout
and stderr from application science codes, and by this mechanism astronomy codes
can send vital information and updates through the events system.
The events are sent to remote Notification services, which gather the events from
distributed processes into a central repository,

A single Elf/OgreScript execution may be composed of a sequence of application modules.
New codes are easily added into processing pipelines by writing simple module descriptions
with information such as the executable to be launched, command line arguments,
and descriptions of input lists or files.

\subsection{Orchestration}

The orchestration layer is written in Perl and utilizes the Perl DBI
module to abstract its interactions with the database persistence
layer. This will enable transparent access to different database types
(Oracle, MySQL, PostgreSQL, etc.).  Perl scripts of the orchestration
layer serve as wrappers for various Condor\footnote{Condor URL:
http://www.cs.wisc.edu/Condor} (Thain, Tannenbaum, \& Livny 2005)
commands, converting between more customized scientific inputs/outputs and
the general Condor inputs/outputs.  For each data-parallel block of codes,
the Orchestration layer must perform several tasks.  First it queries
the central database to get a list of input images and divides that into
sublists for the data-parallel jobs.  It stages the input images to the
target machine.  The orchestration then creates the Elf/OgreScript XML
scripts and properties files for the jobs and stages these files and
the input lists to the target machine.  Then the orchestration submits
the pipeline jobs using vanilla Condor jobs if submitting to the local
condor pool or using Condor-G jobs to remote target machines such as the
TeraGrid cluster.   To automatically control the sequence of these steps
through the entire image processing, the orchestration uses Condor's
Directed Acyclic Graph Manager (DAGMan).

Orchestration uses the Data Access Framework (DAF) to stage files in
multiple cases.  The DAF is a set of programs that can also be used
by operators to transfer files while updating the central database.
There are locally created input lists and files that need to be copied
to the target machine.   Sometimes the images need to be copied from
other archive locations.  And after processing, the newly created images
and files can be backed up to other archive locations.  The processing
framework also uses the centralized database to create input lists of
images and metadata for the target jobs.

\section{Results}

The processing framework has been used to process several pipelines on 
simulated DES data as well as real data from the Blanco Cosmology Survey (BCS). 
An earlier version was used to reduce 45 nights of simulated DECam
supernova imaging data.  In Data Challenge 4 (DC4), which finished at the end
of January, 2009, the processing framework was used to process three different 
astronomy pipelines for almost the entire DC4 data. 
These include the nightly processing pipeline, 
coaddition pipeline and Weak lensing pipeline. The nitely processing pipeline
involves  crosstalk corrections, detrending, astrometric refinement, followed by remapping and catalog ingestion. The
nitely processing was run on 10 nights of simulated DES data, out of which 3
were nonphotometric. The coaddition and weak-lensing pipelines were processed on a 
tile-by-tile basis where the sky was divided into nonoverlapping rectangular 
regions called tiles. The coaddition and weak lensing pipelines were run on  
about 250 tiles for simulated DC4 data. The coaddition pipeline involves combining multiple
images of the same region of sky and different bands into multiple images.
Weak lensing pipeline involves identification of bright stars which are useful
for PSF description, measurement of shapelet decomposition of stars and 
estimation of shear of deconvolved galaxies. In addition, we also developed the PSF 
homogenization and difference imaging pipeline which was tested on a small subset of DC4 
data. More details of these pipelines can be found in Mohr et al (2008). 

\section{Acknowledgements}

The DESDM team acknowledges continuing support from NSF AST 07-15036
and NSF AST 08-13534 as well as significant seed funding provided by
the NCSA, Astronomy, LAS College, and the Vice Chancellor for Research.

The Collaborating Institutions are Argonne National Laboratories,
the University of Cambridge, Centro de Investigaciones Energeticas,
Medioambientales y Tecnologicas-Madrid, the University of Chicago,
University College London, DES-Brazil, Fermilab, the University of
Edinburgh, the University of Illinois at Urbana-Champaign, the Institut de
Ciencies de l'Espai (IEEC/CSIC), the Institut de Fisica d'Altes Energies,
the Lawrence Berkeley National Laboratory, the University of Michigan,
the National Optical Astronomy Observatory, the Ohio State University,
the University of Pennsylvania, the University of Portsmouth and the
University of Sussex.


\begin{references}
\reference Thain, D.\, Tannenbaum, T.\, \& Livny, M. 2005, CONCURRENCY-PRACT EX, 17, 2-4, 323
\reference Noordam, J.~E.\ 2004, Proc.\ SPIE, 5489, 817
\reference Mohr, J.~J\ et al, 2008,  Proc.\ SPIE, 7016, 70160L
\end{references}
\end{document}